\documentclass[a4paper]{aa}

\usepackage{graphicx,float}
\usepackage{latexsym,amsmath,amssymb}
\usepackage{natbib}
\bibpunct{(}{)}{;}{a}{}{,}

\def \xmm {XMM-Newton}
\def \src {GRO\,J1655$-$40}
\def \degmark{^\circ}
\def \nh {N${\rm _H}$}

\def \hcm {\hbox {\ifmmode $ atom cm$^{-2}\else atom cm$^{-2}$\fi}}
\def \arcmin {\hbox{$^\prime$}}
\def \arcsec {\hbox{$^{\prime\prime}$}}
\def \deg {$^{\circ}$}
\def \chisq {$\chi ^{2}$}
\def \rchisq {$\chi_{\nu} ^{2}$}
\def\approxgt{\mathrel{\hbox{\rlap{\lower.55ex \hbox {$\sim$}}
        \kern-.3em \raise.4ex \hbox{$>$}}}}
\def\approxlt{\mathrel{\hbox{\rlap{\lower.55ex \hbox {$\sim$}}
        \kern-.3em \raise.4ex \hbox{$<$}}}}

\newcommand {\Msun}{M_\odot}

\newcommand {\phind} {$\Gamma$}
\newcommand {\oone} {\ion{O}{i}}
\newcommand {\otwo} {\ion{O}{ii}}

\newcommand {\fetfour} {\ion{Fe}{xxiv}}
\newcommand {\fetfive} {\ion{Fe}{xxv}}
\newcommand {\fetsix} {\ion{Fe}{xxvi}}
\newcommand {\ka} {K$\alpha$}
\newcommand {\kb} {K$\beta$}

\newcommand {\fetthree} {\ion{Fe}{xxiii}}

\newcommand {\oseven} {\ion{O}{vii}}
\newcommand {\oeight} {\ion{O}{viii}}

\newcommand {\netwo} {\ion{Ne}{ii}}
\newcommand {\nenine} {\ion{Ne}{ix}}
\newcommand {\neten} {\ion{Ne}{x}}

\newcommand {\nitseven} {\ion{Ni}{xxvii}}
\newcommand {\niteight} {\ion{Ni}{xxviii}}

\def \mxb {MXB\,1658$-$298}

\def \grs {GRS\,1915+105}
\def \gro {GRO\,J1655$-$40}
\def \gx {GX\,13+1}

\def \seventeen {H\,1743$-$322}
\def \threethreenine {GX\,339$-$4}
\def \xte {XTE\,J1650$-$500}

\def \nhxabs {$N{\rm _H^{xabs}}$}
\def \nhwarmabs {$N{\rm _H^{warmabs}}$}

\def \xiunit {\hbox{erg cm s$^{-1}$}}
\def \logxi {$\log(\xi)$}

\newcommand {\egau} {$E_{\rm gau}$}
\newcommand {\eedge} {$E_{\rm edge}$}
\newcommand {\ktbb} {$kT_{\rm bb}$}

\newcommand {\sig} {$\sigma$}
\newcommand {\ew} {$EW$}
\newcommand {\ews} {$EW$s}
\newcommand {\lya} {Ly$\alpha$}
\newcommand {\lyb} {Ly$\beta$}

\def \nh {$N{\rm _H}$}
\def \nhabs {$N{\rm _H^{abs}}$}
\def \nhxabs {$N{\rm _H^{xabs}}$}

\newcommand {\ttnh} {$\times~$10$^{22}$~atom~cm$^{-2}$}

\def \xiunit {\hbox{erg cm s$^{-1}$}}
\def \logxi {$\log(\xi)$}
\def \xil {$\xi$}

\def \warmabs {{\tt warmabs}}
\newcommand {\sigmav} {$\sigma_{\rm v}$}
\newcommand {\kms} {km~s$^{-1}$}

\def \kpl {$k_{\rm pl}$}
\def \kbb {$k_{\rm bb}$}
\def \kgau {$k_{\rm gau}$}

\begin{document}

\title{XMM-Newton and INTEGRAL spectroscopy of the microquasar GRO\,J1655$-$40 during its 2005 outburst}

\author{M. D{\'i}az Trigo\inst{1} \and A. N. Parmar\inst{1}
        \and J. Miller\inst{2} \and E. Kuulkers\inst{3} \and
	M. D. Caballero-Garc{\'i}a\inst{4}
        }

\institute{
       Astrophysics Mission Division, Research and Scientific Support
       Department of ESA, ESTEC,
       Postbus 299, NL-2200 AG Noordwijk, The Netherlands
\and
       Department of Astronomy, University of Michigan, 500 Church Street, Dennison 814, Ann Arbor, MI 48109, USA
\and
       Integral Science Operations Centre, Science Operations and Data Systems Division,
       Research and Scientific Support Department of ESA, ESAC, Apartado 50727, 28080
       Madrid, Spain
\and
       Laboratorio de Astrof{\'i}sica Espacial y F{\'i}sica Fundamental, INTA, Apartado 50727,
       28080 Madrid, Spain
}

\date{Received ; Accepted:}

\authorrunning{D{\'i}az Trigo et al.}

\titlerunning{XMM-Newton and INTEGRAL observations of \gro}

\abstract{We report on two simultaneous XMM-Newton and INTEGRAL
observations of the microquasar \gro\ during the 2005 outburst
when the source was in its high-soft state. The 0.3--200~keV
spectra are complex with an overall continuum which may be modeled
using an absorbed blackbody together with a weak, steep, power-law
component. In addition, there is evidence for the presence of {\it
both} a relativistically broadened Fe K line and a highly
photo-ionized absorber. The photo-ionized absorber is responsible
for strong K absorption lines of \fetfive\ and \fetsix\ in the
EPIC pn spectra and blue-shifted ($v = -540 \pm 120$~\kms) \neten\
and \fetfour\ features in the RGS spectra. The parameters of the
photo-ionized absorber were different during the two observations
with the ionization parameter, \logxi, decreasing from $3.60 \pm
0.04$ to $3.30 \pm 0.04$~\xiunit\ and the column density
decreasing from $(5.2 \pm 1.0)$~\ttnh\ to $(1.5 \pm 1.0)$~\ttnh\
during the first and second observations as the 0.5--200~keV \gro\
luminosity decreased by around a half. At 90\% confidence the
INTEGRAL data show no evidence of a break in the power-law
component up to energies of 380~keV and 90~keV for the first
and second observations, respectively. }

\maketitle

\section{Introduction}
\label{sect:intro}

%%----------------------------------------------------------------------

The X-ray nova \gro\ was discovered using the Burst and Transient
Source Experiment (BATSE) onboard the Compton Gamma Ray
Observatory \citep{1655:zhang94iauc}. Soon after the discovery
radio observations revealed apparently superluminal relativistic
jets moving in opposite directions almost perpendicular to
the line of sight (85\deg) with a velocity of 0.92~c
\citep{1655:hjellming95nature, 1655:tingay95nature}. Optical
observations, obtained when the system had returned to quiescence,
showed that the companion star is a F3--5 giant or sub-giant in a
2.62~day orbit around a 5--8~$\Msun$ compact object, which is
almost certainly a black hole \citep{1655:orosz97apj,
1655:vanderhooft98aa, 1655:shahbaz99mnras}. \gro\ has been
observed to undergo deep absorption dips
\citep{1655:kuulkers98apj, 1655:balucinska01adspr,
1655:tanaka03mnras, 1655:kuulkers00aap} consistent with observing the source at an
inclination angle of $\sim 70\degmark$
\citep{1655:vanderhooft98aa}. This indicates that the inclination
of the inner disk may differ slightly from the inclination of the
binary system. This misalignment may arise if the jet axis
coincides with the spin axis of the compact object and the inner
disk is anchored perpendicular to this axis by gravitational and
magneto-hydrodynamic effects. The misalignment of the binary plane
and the jet could be explained by the Bardeen-Petterson effect
\citep{maccarone02mnras}.
\gro\ has been subject of a considerable number of studies using
X-ray timing, presenting the highest-frequency quasi-periodic
oscillations (QPOs) yet seen in a black hole \citep[450
Hz,][]{1655:strohmayer01apjl}.

Narrow X-ray absorption lines from highly ionized Fe were first
detected from \gro\ \citep{1655:ueda98apj, 1655:yamaoka01pasj} and
from another superluminal jet source \grs\
\citep{1915:kotani00apj,1915:lee02apj}. Similar features were
subsequently detected from the low-mass X-ray binary (LMXB) \gx\
\citep{gx13:ueda01apjl} and from all the bright dipping LMXBs
observed with XMM-Newton \citep{ionabs:diaz06aa}. The changes in
the X-ray continuum and the Fe absorption features during dips are
explained as resulting primarily from an increase in column
density and a decrease in the ionization state of a highly-ionized
absorber \citep{1323:boirin05aa,ionabs:diaz06aa}. Since dipping
sources are simply normal LMXBs viewed from close to the orbital
plane, this implies that ionized absorbers are a common feature of
LMXBs. Outside of the dips, the properties of the absorption
features do not vary strongly with orbital phase. This suggests
that the ionized plasma in LMXBs has a cylindrical geometry with a
maximum column density close to the plane of the accretion disk.

Recently {\it Chandra} HETGS observations of the black hole
candidates \threethreenine, \xte\ and \seventeen\
\citep{gx339:miller04apj, 1743:miller04apj} have revealed the
presence of variable, blue-shifted, highly-ionized absorption
features which are interpreted as evidence for outflows. While
\fetfive\ and \fetsix\ features are present in the \seventeen\
spectrum, \threethreenine\ and \xte\ show \oeight\ and \nenine\ or
\netwo\ features from less ionized material. These features
suggest that a warm absorber analogous to those seen in many
Seyfert galaxies is present in systems such as \threethreenine\
and \xte\ \citep{gx339:miller04apj}. In contrast,
\citet{1743:miller04apj} propose that the absorber present in
\seventeen\ is more highly ionized.

\begin{figure}[!ht]
\centerline{\hspace{-1cm}\includegraphics[angle=0,width=9.0cm,height=6cm]{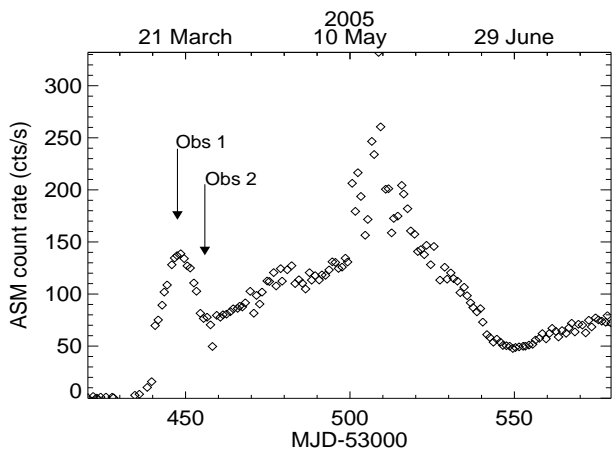}}
\caption{ASM 1.5--12 keV lightcurve of \src. The times of the
XMM-Newton and INTEGRAL observations reported here are indicated
with arrows.} \label{fig:lc-rxte}
\end{figure}

We report on two simultaneous XMM-Newton and INTEGRAL observations
of \gro\ obtained during a recent outburst that started around
2005 February 17 \citep{1655:markwardt05atel} and reached a
maximum intensity of $\sim$5 Crab during flares around 2005 May
20. RXTE \citep{1655:homan05atel} and {\it Swift} \citep{1655:brocksopp06mnras}
monitored the outburst and near-simultaneous
observations to the \xmm\ and INTEGRAL observations are available. 
Fig.~\ref{fig:lc-rxte} shows the 
RXTE All-Sky Monitor (ASM) lightcurve of part of the 2005
outburst with the times of the XMM-Newton and INTEGRAL
observations indicated.

\section{Observations and data analysis}
\label{sec:reduction}

\subsection{XMM-Newton observations}
The XMM-Newton Observatory \citep{jansen01aa} includes three
1500~cm$^2$ X-ray telescopes each with an European Photon Imaging
Camera (EPIC, 0.1--15~keV) at the focus.  Two of the EPIC imaging
spectrometers use MOS CCDs \citep{turner01aa} and one uses pn CCDs
\citep{struder01aa}. Reflection Grating Spectrometers \citep[RGS,
0.35--2.5~keV,][]{denherder01aa} are located behind two of the
telescopes. \src\ was observed by XMM-Newton for 23.8~ks on 2005
March 18 between 15:47 and 22:25~UTC (Obs~ID~0155762501,
hereafter called Obs~1) and for 22.4~ks on 2005 March 27 between
08:43 and 14:56~UTC (Obs~ID~0155762601, hereafter called
Obs~2). The thin optical blocking filter was used with the EPIC
camera. The EPIC pn camera was operated in burst mode due to the
high count rate of the source ($\approxgt$5000~s$^{-1}$). For the
same reason, the EPIC MOS cameras were not operated. RGS2 was
operated in its normal spectroscopy mode with all CCD chips read
in parallel. However, the individual RGS1 CCD chips were read out
sequentially. This reduces the frame time from 4.8 to 0.6 seconds
(4.8/8) and consequently the pile-up by almost an order of
magnitude. All the X-ray data products were obtained from the
XMM-Newton public archive and reduced using the Science Analysis
Software (SAS) version 6.1.0.

In pn burst mode, only one CCD chip (corresponding to a field of
view of 13\farcm6$\times$4\farcm4) is used and the data from that
chip are collapsed into a one-dimensional row (4\farcm4) to be
read out at high speed, the second dimension being replaced by
timing information. The duty cycle is only 3\%. This allows a time
resolution of 7~$\mu$s, and photon pile-up occurs only for count
rates above 60000~s$^{-1}$. Only single and double events
(patterns 0 to 4) were selected. Source events were extracted from
a 53\arcsec\ wide column centered on the source position (RAWX 30
to 43). Background events were obtained from a column of the same
width, but centered 115\arcsec\ from \src\ (RAWX 2 to 15). Due to
the brightness of the source, the region used to extract the
background is highly contaminated by source events. For this
reason we did not subtract the background from the source which is
anyway negligible for such a strong source. Ancillary response
files were generated using the SAS task {\tt arfgen}. Response
matrices were generated using the SAS task {\tt rmfgen}. We used
burst mode pn data in the 0.5--10~keV energy range.

The SAS task {\tt rgsproc} was used to produce calibrated RGS
event lists, spectra, and response matrices. The RGS2 spectra were
strongly piled-up and could not be used reliably. Pile-up in the
XMM-Newton gratings is rare, and we have been unable to find a
method to correct for pile-up in the literature. Less than
5\% of events are affected by pile-up in all CCDs except number 8
covering the energy range 1.18--1.6~keV, 
where pile-up may have reached
25\% during Obs~1 and 15\% in Obs~2. This will strongly affect
continuum measurements, but the derived properties of any narrow
line features will not be so strongly affected. Therefore we only
used RGS1 data from the first and second orders.

\begin{figure*}[!ht]
\centerline{\hspace{-1cm}\includegraphics[angle=0]{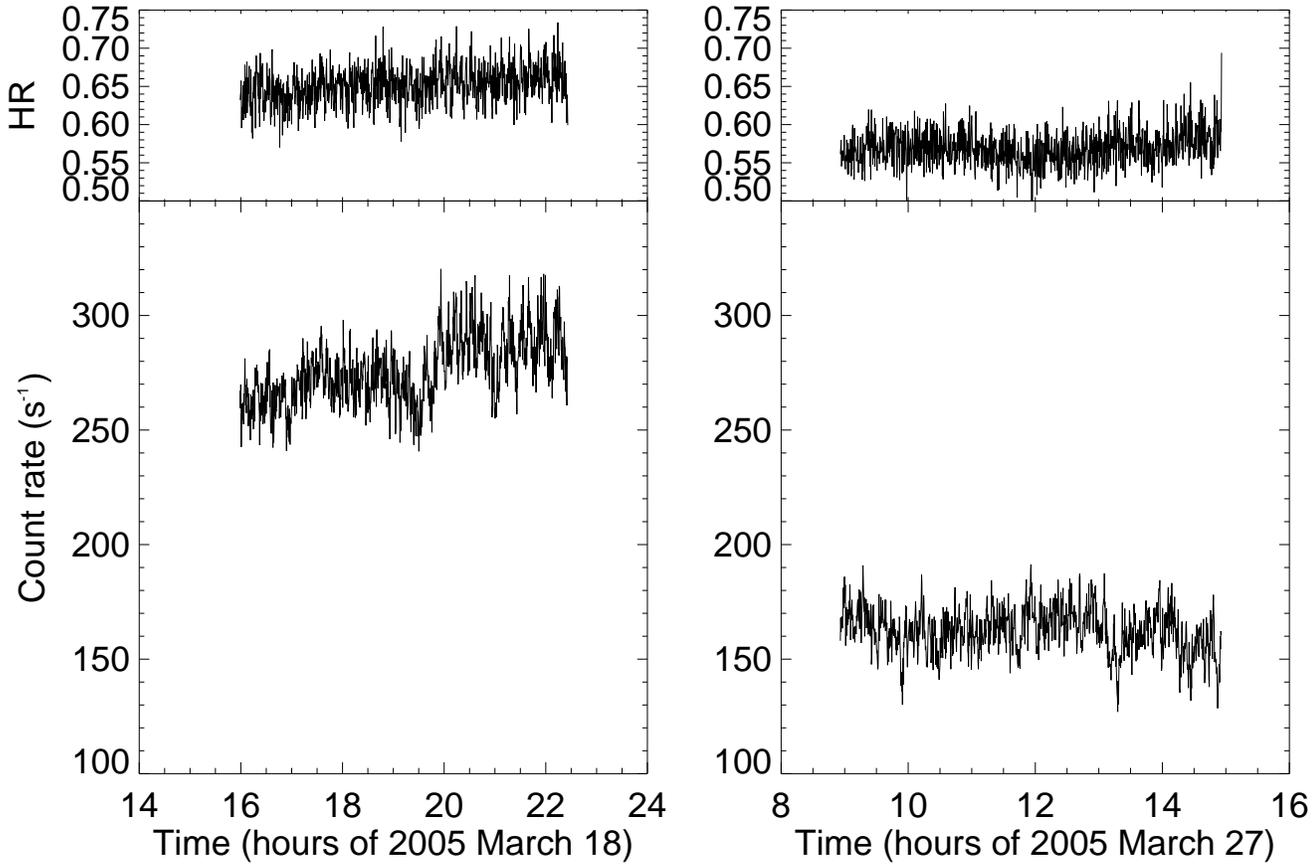}}
\caption{Lower panels: 0.6--10~keV EPIC pn lightcurve of \src\ for
Obs~1 (left panel) and Obs~2 (right panel). The count rate
has not been corrected for the 3\% duty cycle of the burst mode.
Upper panels: Hardness ratio (counts in the 2.5--10~keV band
divided by those between 0.6--2.5~keV) for Obs~1 (left panel) and
Obs~2 (right panel). The binning is 20~s for all panels.}
\label{fig:lc}
\end{figure*}

\subsection{INTEGRAL observations}

The INTEGRAL \citep{Winkler03AA} payload consists of two main
gamma-ray instruments, one of which is optimized for 15~keV to 10~MeV
high-resolution imaging \citep[IBIS;][]{Ubertini03AA} and the other
for 18~keV to 8~MeV high-resolution spectroscopy
\citep[SPI;][]{vedrenne03AA}. IBIS provides an angular resolution of
$12\arcmin$ full-width at half-maximum (FWHM) and an energy
resolution, $E/\Delta E$, of $\sim$12~FWHM at 100~keV. SPI provides an
angular resolution of $2\fdg 5$~FWHM and an $E/\Delta E$ of
$\sim$430~FWHM at 1.3~MeV. The extremely broad energy range of IBIS is
covered by two separate detector arrays, ISGRI
\citep[15--1000~keV;][]{lebrun03AA} and PICsIT
\citep[0.175--10~MeV;][]{labanti03AA}. The payload is completed by
X-ray \citep[JEM-X; 3--35~keV;][]{Lund03AA} and optical monitors
\citep[OMC; V-band;][]{Mas-Hesse03AA}.  JEM-X has a fully coded FOV of
$4\fdg8$ diameter and an angular resolution of $3\arcmin$ FWHM.  All
the instruments are co-aligned and are operated simultaneously.

As part of a Target of Opportunity programme on known black hole
candidates, two INTEGRAL observations of the region of sky
containing \src\ were performed between 2005 March 18 07:42 and
2005 March 19 04:22~UTC, for a total time of 70~ks and between
2005 March 26 19:00 and 2005 March 28 00:56~UTC, for a total time
of 100~ks, coinciding with the above mentioned \xmm\ observations.
The standard 5 x 5 dither patterns of pointings centered on the
target were performed.

Data from the two observations were processed using the Off-line
Scientific Analysis (OSA) version 5.1 software provided by the
INTEGRAL Science Data Centre \citep[ISDC;][]{Courvoisier03AA}.
This includes pipelines for the reduction of INTEGRAL data from
all four instruments. The three high-energy instruments use coded
masks to provide imaging information. This means that photons from
a source within the field of view (FOV) are distributed over the
detector area in a pattern determined by the position of the
source in the FOV. Source positions and intensities are determined
by matching the observed distribution of counts with those
produced by the mask modulation. In this paper we use data from
JEM-X and ISGRI.

\section{Results}

\subsection{Soft X-ray lightcurve}

The EPIC pn 0.6--10~keV lightcurves from both observations are
shown in Fig.~\ref{fig:lc} (lower panels) with a binning of 20~s.
The upper panels show the hardness ratio (counts in the
2.5--10~keV energy range divided by those between 0.6--2.5~keV)
also with a binning of 20~s. The source intensity increased from
$\sim$260~count~s$^{-1}$ to $\sim$300~count~s$^{-1}$ during
Obs~1. This increase is associated with an increase of the
hardness ratio from 0.63 to 0.67, indicating that the source has not yet reached the
outburst peak. In contrast, Obs~2 shows a lower count rate of
$\sim$160~s$^{-1}$ with no obvious trend present and a nearly
constant hardness ratio. Strong variability is present during both
observations. Although we do not plot the error bars in
Fig.~\ref{fig:lc} for clarity, they are small compared to the
average deviations seen. Two 10\% and one 15\% irregular
reductions in intensity are present in Obs~1 and Obs~2,
respectively.

\subsection{X-ray spectra}
\label{sec:averaged}

We performed spectral analysis using XSPEC \citep{arnaud96conf}
version 12.3.0. We
used the photo-electric cross-sections of \citet{wilms00apj} to
account for absorption by neutral gas with solar abundances
({\tt tbabs} model in XSPEC). Spectral
uncertainties are given at 90\% confidence ($\Delta$\chisq = 2.71
for one parameter of interest), and upper limits at 95\%
confidence.
We quote all line equivalent widths ($EW$s) with positive values
for both absorption and emission features. We rebinned all spectra
to over-sample the $FWHM$ of the energy resolution by a
factor 3(5) for EPIC(RGS), and to have a minimum of 25 counts per
bin, to allow the use of the $\chi^2$ statistic. The systematic
effects in the EPIC pn burst mode have been estimated as $\sim$5\%
(M. Kirsch, private communication). These effects are likely to be
localized at low energies, mainly below 3~keV, and especially in
the 1.5--3~keV energy band, where sharp drops at $\sim$1.9, 2.3
and 2.7~keV are present in the pn effective area. To account for
such effects, we added quadratically a 5\% uncertainty to
each pn spectral bin in the 1.5-3~keV energy range and 2\% in the
rest of the energy  band. We added 2\% systematic
uncertainty to the JEM-X and ISGRI spectra. In the following we
present the results of fits to the RGS, EPIC pn, JEM-X and
ISGRI spectra of \gro\ during the 2005 outburst.
\footnote[1] {We note that for Obs~2 we have removed one bin between 20 and 30 keV
which does not change the results of the fit but increases
significantly the \rchisq. This bin may indicate a calibration feature or
may have astrophysical origin. However, it is beyond the scope of this
paper to analyse its nature and this will be done in a future paper about
INTEGRAL data (Caballero-Garc{\'i}a et al., in preparation).}

\subsubsection{Joint EPIC pn, JEM-X and ISGRI spectral analysis}
\label{sec:pn-isgri}

We first fit the combined 0.5--10~keV EPIC pn, 5--20~keV JEM-X
and 20--200~keV ISGRI spectra with a model consisting of a
disk-blackbody and a power law, both modified by photo-electric
absorption from neutral material (model {\tt tbabs*(dbb+po)}).
Constant factors, fixed to 1 for the EPIC pn spectra, but allowed
to vary for the JEM-X and ISGRI spectra, were included
multiplicatively in order to account for cross-calibration
uncertainties. Previous RXTE and XMM-Newton spectra of \src\
were fit with a multi-color disk-blackbody plus a power-law model
with, or without, a cutoff
\citep[e.g.,][]{1655:ueda98apj,1655:kuulkers98apj,1655:tanaka03mnras}.
We note that, while a blackbody is a better fit to the low-energy
part of the spectrum (0.5--10~keV), a disk-blackbody gives
significantly better fits when the INTEGRAL 5--200~keV data
are included. This is due to the fact that the high-energy data
constrain the power-law component well, and consequently
allow the other continuum components to be better studied.

The fits to the combined spectra with the {\tt
tbabs*(diskbb+po)} model are unacceptable with \rchisq\ of
9.3 and 19.0 for 293 and 256 degrees of freedom (d.o.f.), for
Obs~1 and Obs~2, respectively. The poor fit quality is mainly
due to the presence of strong absorption features near 7~keV and
strong emission features in the pn below 3 keV. A broad emission
feature is superposed on the 7~keV absorption features. Structured
residuals near 2.3~keV and 2.8~keV are probably due to an
incorrect instrumental modeling of the Au mirror edges, whilst
those near 1.8~keV are probably due to an incorrect modeling of
the Si absorption in the CCD detectors. In addition, at the
energies of these edges there is a sharp drop in the effective
area of the pn. Both effects do not appear to be well accounted
for in the EPIC pn calibration. An excess in the pn spectra is
detected at $\sim$0.6~keV and can be modeled by a Gaussian
emission line. This feature is present in both observations and
could be emission from \oseven. In addition, an absorption feature
in the pn spectra is detected at $\sim$0.5~keV. Several absorption
features are detected in the RGS at this energy and are attributed
to \oone\ and \otwo\ (see Sect.~\ref{sect:rgs}). Finally, an
excess near 1~keV is detected in the EPIC pn spectrum in both
observations. This feature is detected in a number of X-ray
binaries and has been previously modeled either as an emission
line, or as an edge, and its nature is unclear
\citep[e.g.,][]{cygx2:kuulkers97aa, 1658:sidoli01aa,
1254:boirin03aa, 1916:boirin04aa,1323:boirin05aa}. If the feature
has an instrumental origin it might be expected to be stronger in
the most luminous observation. The fact that the feature is very
strong in the less luminous observation and only a weak excess is
observed in Obs~1 could indicate that it has an astrophysical
origin and its energy is consistent with a blend of \nenine\ and
\neten\ emission, or Fe L emission.

We included a narrow Gaussian emission feature at 2.3~keV
({\tt gau$_5$}) to account for the burst mode calibration
deficiencies discussed previously. A Gaussian absorption line {\tt
gau$_1$} was added to account for the absorption at $\sim$0.5~keV.
The excesses at $\sim$0.6~keV and $\sim$1~keV were modeled by
three Gaussian emission lines, {\tt gau$_{2-4}$}. The first
Gaussian accounts for the excess at 0.6~keV. The other two
Gaussians account for the excess at 1~keV and their energies were
fixed at 0.922 and 1.022~keV, the energies of \nenine\ and \neten\
resonance transitions. These additions reduce the \rchisq\ to 8.9
for 284 d.o.f for Obs~1 and to 8.7 for 247 d.o.f. for Obs~2.

To account for the complex residuals near 7~keV, absorption from a
photo-ionized plasma ({\tt warmabs} model within XSPEC) and a
relativistic emission line ({\tt laor}) modified by neutral
absorption were added to the model. The {\tt laor} model
\citep{laor91apj} describes the line profile expected when an
accretion disk orbiting a black hole with maximum angular momentum
is irradiated by a source of hard X-rays. The \rchisq\ of the fit
is significantly reduced to 1.5 for 276 d.o.f. and 2.1 for
239 d.o.f. for Obs~1 and Obs~2, respectively.

We substituted the relativistic emission line by a broad Gaussian
emission line to check whether such a line may be broadened
by effects other than relativity. The \rchisq\ of the fit
increases, compared to the previous fit, to 9.3 for 278
d.o.f. for Obs~1 and to 9.3 for 241 d.o.f. for Obs~2. Thus, we
conclude that relativistic broadening is more favoured than e.g.
Compton scattering or the contribution from Fe with a range of
ionization states for both observations. Finally we
substituted the {\tt laor} component by the {\tt laor2} model in
XSPEC, which uses a broken power-law emissivity profile instead of
a power-law. The emissivity profile describes the dependence of
the emissivity with radial position on the disk. This change
reduces the \rchisq\ to 1.27 for 274 d.o.f. for Obs~1 and to 1.85
for 237 d.o.f. for Obs~2.

\begin{figure*}[ht!]
\centerline{\includegraphics[width=0.6\textwidth]{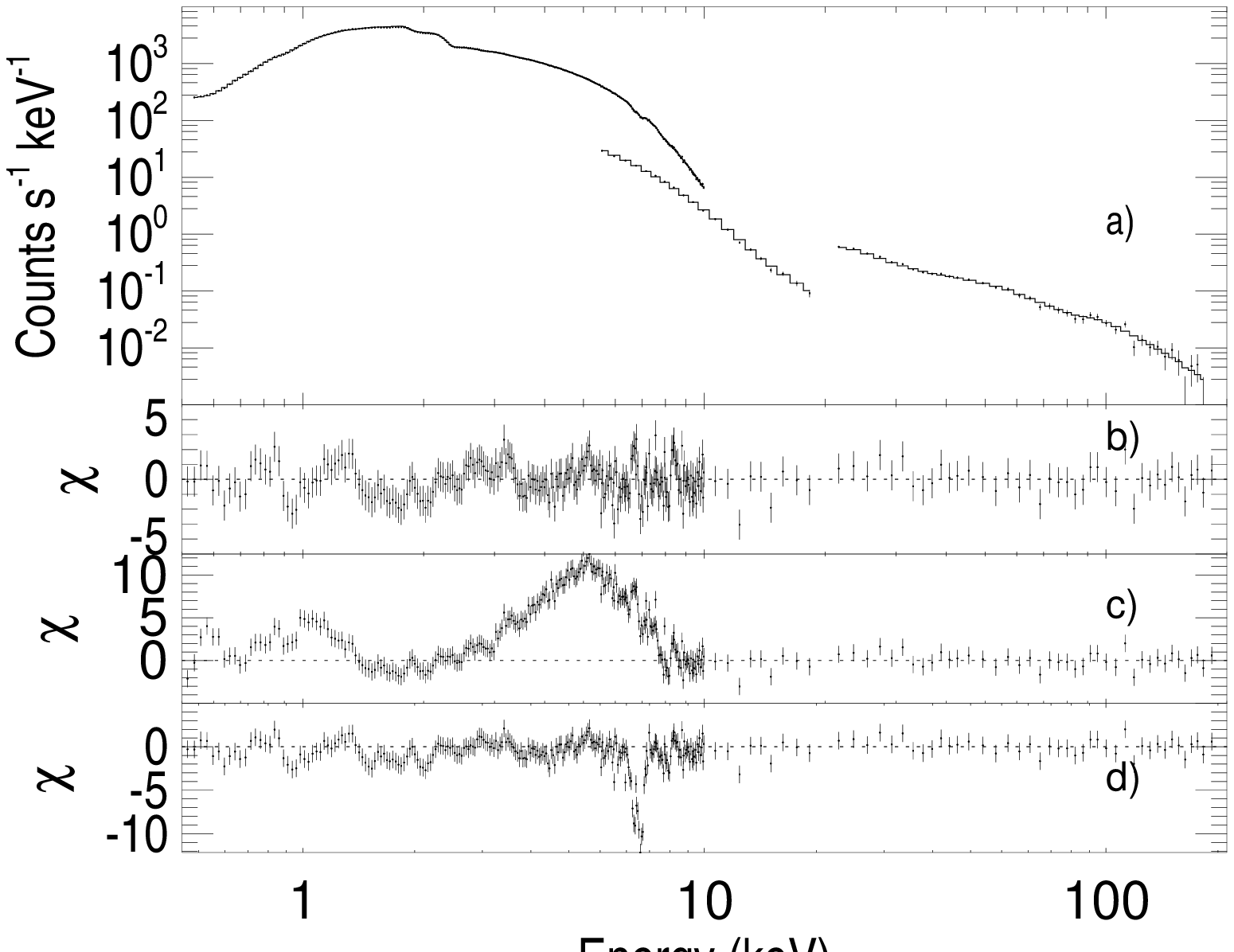}
\hspace{-1.1cm}
\includegraphics[width=0.6\textwidth]{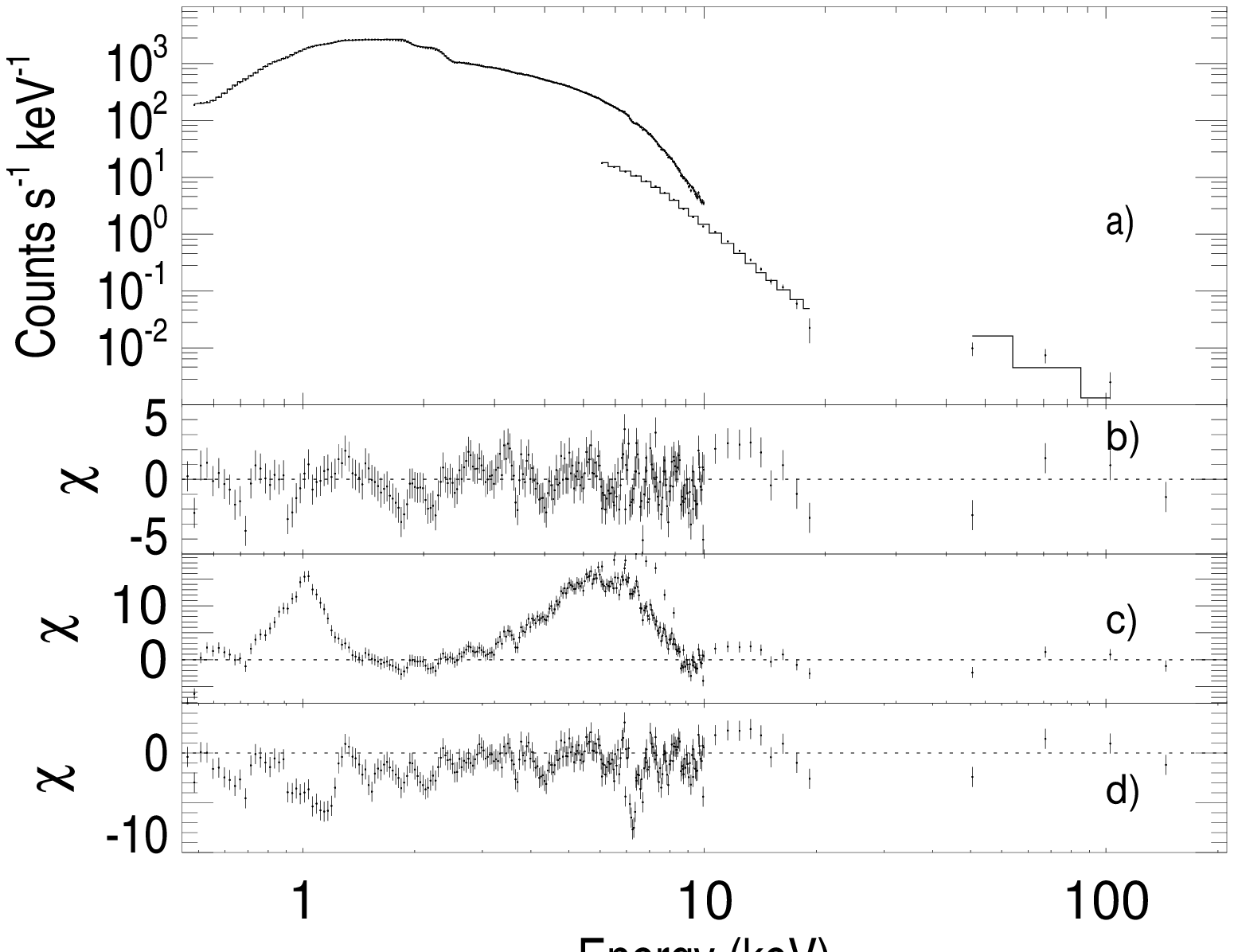} }
\caption{ (a) \src\ EPIC pn, JEM-X and ISGRI spectra from Obs~1 (left)
and Obs~2 (right) fit with a model consisting of disk-blackbody
(dbb) and power-law (po) components modified by absorption from
neutral (tbabs) and ionized (warmabs) material together with a
relativistic emission line (laor2), four emission
Gaussians (gau$_{1-4}$) modified by absorption from neutral material (tbabs) and a
Gaussian (gau$_5$) to account for calibration uncertainties. (b)
Residuals in units of standard deviation from the above model. (c)
Residuals when the normalisations of the laor2 component,
and the Gaussians are set to 0. The
laor2 component indicates relativistic broadening. (d)
Residuals when \nhwarmabs\ is set to 0. Absorption lines are clearly
visible near 7~keV. The best-fit parameters are given in
Table~\ref{tab:bestfit}.} \label{fig:spectrum}
\end{figure*}

\begin{figure*}[ht!]
\centerline{\includegraphics[width=0.6\textwidth]{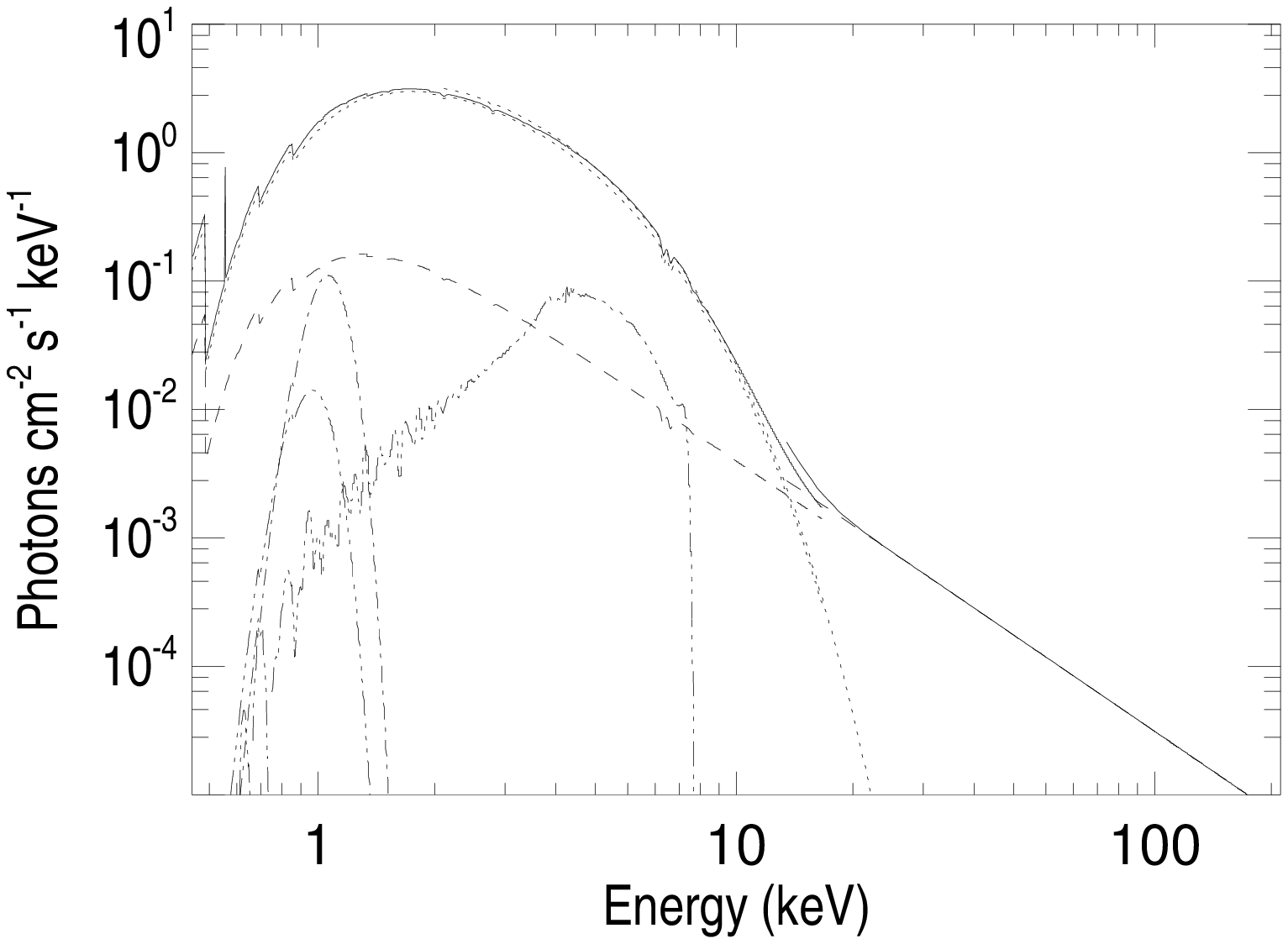}
\hspace{-1.1cm}
\includegraphics[width=0.6\textwidth]{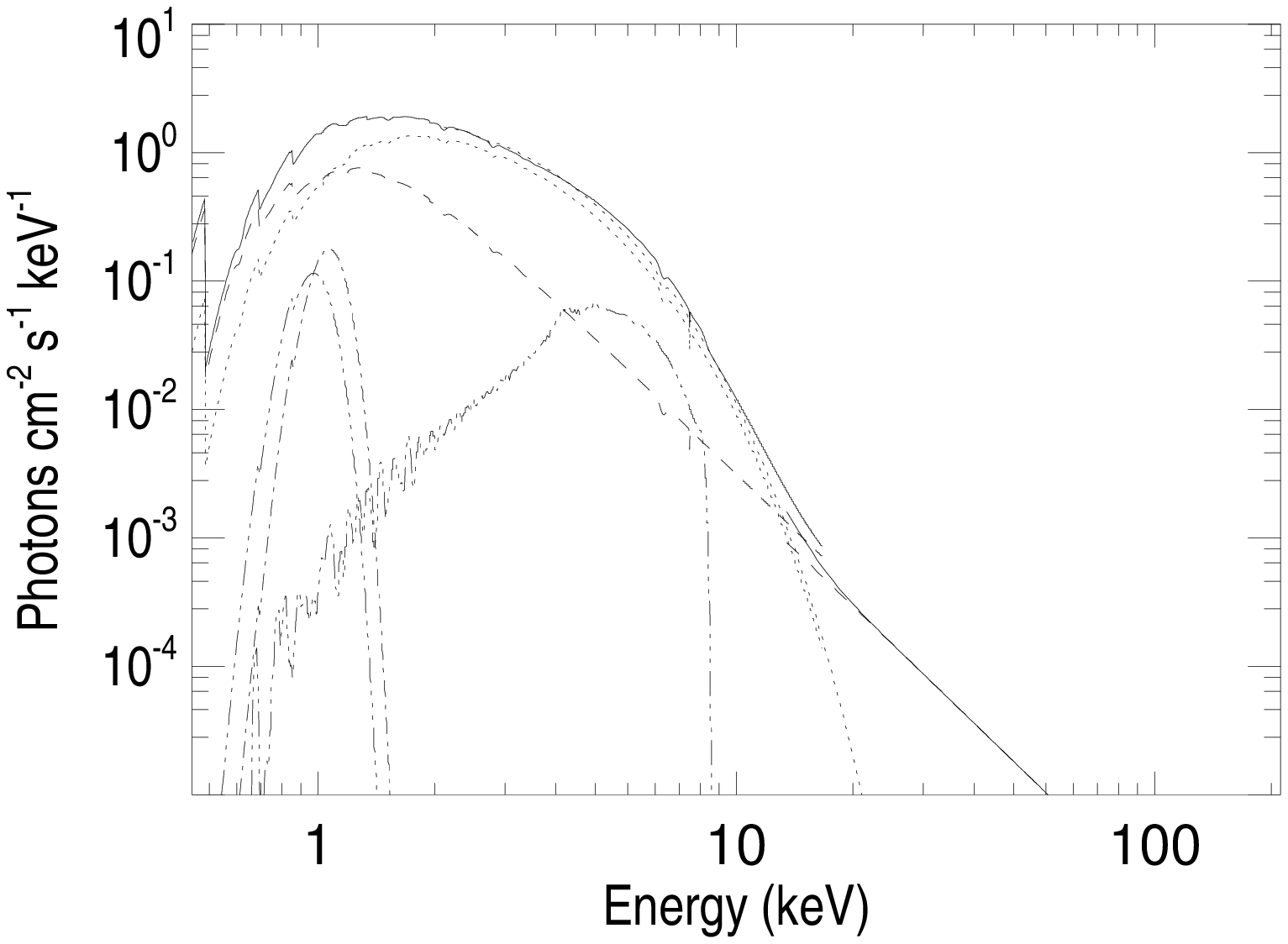} }
\caption{ \gro\ photon spectrum during Obs~1 (left panel) and
Obs~2 (right panel) with the separate components indicated:
disk-blackbody (dotted line), power-law (dashed line), laor2
relativistic line (dot-dashed line) and soft excess at 1~keV
(three dots-dashed line). The best-fit parameters are given in
Table~\ref{tab:bestfit}.} \label{fig:model}
\end{figure*}

The parameters of the best-fit model (model~1: {\tt
tbabs*warmabs*(diskbb+po) +
tbabs*(laor2+gau$_1$+gau$_2$+gau$_3$+gau$_4$) + gau$_5$}) are
given in Table~\ref{tab:bestfit}. Fig.~\ref{fig:spectrum} shows
the best-fit model and residuals and Fig.~\ref{fig:model} the
components of the model. We have checked for the presence of
a cutoff in the power-law component by including a high-energy
cutoff and calculating the 90\% confidence limits. We find no
evidence of a break in the power-law component up to energies of
380~keV and 90~keV for Obs~1 and Obs~2, respectively.

\begin{table}[]
\begin{center}
\caption[]{
  Best-fits to the 0.5--10~keV EPIC pn, 5--20~keV JEM-X and 20--200~keV ISGRI spectra using the
  {\tt tbabs*warmabs*(diskbb+po)+ tbabs*(laor2+gau$_1$+gau$_2$+gau$_3$+gau$_4$)+ gau$_5$ model.}
  \kpl, \kbb\ and \kgau\ are the normalizations of the power law, disk-blackbody and
  Gaussian emission
  lines, respectively. L$_{0.5-200~keV}$ is the 0.5--200 keV luminosity between for
 a distance of 3.2~kpc. The widths of the Gaussian emission lines gau$_{2-4}$ are constrained
  to be $\le$0.1~keV.
}
\begin{tabular}{lc@{\extracolsep{0.15cm}}c@{\extracolsep{0.15cm}}c@{\extracolsep{0.15cm}}}
\hline
%\noalign{\smallskip}
\hline
%\noalign{\smallskip}
 & & Obs 1 & Obs 2 \\
% & & Fit 1 & Fit 2 & Fit 1 & Fit 2 \\
\noalign{\smallskip\hrule\smallskip}
%& Comp. & & \\
Parameter & & &\\
& {\tt po} & & \\
\phind & & 2.23 $\pm$ 0.02 &  3.14 $\pm$ 0.02 \\
\multicolumn{2}{l}{\kpl\ } & 0.69 $\pm$ 0.04 & 4.5 $\pm$ 0.2\\
\multicolumn{3}{l}{\small [ph keV$^{-1}$ cm$^{-2}$ s$^{-1}$ at 1 keV]}\\
& {\tt diskbb} & &\\
\ktbb\ {\small[keV]} & & 1.320 $\pm$ 0.003 & 1.303 $\pm$ 0.003 \\
\kbb\   & & 849 $\pm$ 5 & 434 $\, ^{+6}_{-3}$ \\
\multicolumn{3}{l}{\small[(R$_{in}$ [km]/(D$_{10}$ [kpc]))$^2$ * cos$\theta$]} \\
& {\tt tbabs} & & \\
\multicolumn{2}{l}{\nhabs\ {\small[$10^{22}$ cm$^{-2}$]}} & 0.665$\, ^{+0.002}_{-0.003}$  & 0.77 $\, ^{+0.02}_{-0.01}$ \\
& {\tt warmabs} & &  \\
\multicolumn{2}{l}{\nhwarmabs\ {\small[$10^{22}$ cm$^{-2}$]}} & 5.2 $\pm$ 1.0  & 1.5 $\pm$ 1.2 \\
\logxi\ {\small[\xiunit]} & & 3.60 $\pm$ 0.04 & 3.30 $\pm$ 0.04 \\
\sigmav\ {\small[km s$^{-1}$]} & & 3500 $\pm$ 900 & 5900 $\pm$ 1200 \\
%Redshift & & 0.005 & \\
\noalign {\smallskip}
& {\tt laor2}  & & \\
\multicolumn{2}{l}{ \egau\ {\small[keV]}}               & 6.60 $\pm$ 0.06 & 6.80$\, ^{+0.03}_{-0.05}$    \\
\multicolumn{2}{l}{ \kgau\ {\small[ph cm $^{-2}$ s$^{-1}$]}} & 0.25 $\pm$ 0.02 & 0.226 $\pm$ 0.006   \\
\multicolumn{2}{l}{ $q_1$ }               & 8.9 $\pm$ 0.2 & 10$\, _{-0.1}$ \\
\multicolumn{2}{l}{ r$_{in}$ {\small[GM/c$^2$]}}               & 1.53 $\pm$ 0.02 & 1.37 $\pm$ 0.01\\
\multicolumn{2}{l}{ $i$ {\small[deg.]}} &  52 $\pm$ 1 & 63.2 $\pm$ 0.2 \\
\multicolumn{2}{l}{ $q_2$ }               & 3.5 $\, ^{+0.5}_{-0.3}$ & 3.0 $\pm$ 0.3\\
\multicolumn{2}{l}{ r$_{break}$ {\small[GM/c$^2$]}} & 3.6 $\pm$ 0.2 & 3.10 $\pm$ 0.09\\
\noalign {\smallskip}
& {\tt gau$_1$}  & & \\
\multicolumn{2}{l}{ \egau\ {\small[keV]}}               & 0.50 (f) & 0.50 (f) \\
\multicolumn{2}{l}{ $\sigma$ {\small[keV]}}               &  0 (f) & 0 (f) \\
\multicolumn{2}{l}{ \kgau\ {\small[ph cm $^{-2}$ s$^{-1}$]}} & 0.4 $\pm$ 0.2 & 2.0 $\pm$ 0.3  \\
\noalign {\smallskip}
& {\tt gau$_2$}  & & \\
\multicolumn{2}{l}{ \egau\ {\small[keV]}}               & 0.60 (f) & 0.60 (f) \\
\multicolumn{2}{l}{ $\sigma$ {\small[keV]}}               & $<$0.02 & $<$0.04 \\
\multicolumn{2}{l}{ \kgau\ {\small[ph cm $^{-2}$ s$^{-1}$]}} & 0.3 $\pm$ 0.1 & $<$0.4 \\
\noalign {\smallskip}
& {\tt gau$_3$}  & & \\
\multicolumn{2}{l}{ \egau\ {\small[keV]}}               & 0.922 (f) & 0.922 (f)  \\
\multicolumn{2}{l}{ $\sigma$ {\small[keV]}}               &  0.1$^{}_{}$ & 0.1 $\, _{-0.01}$ \\
\multicolumn{2}{l}{ \kgau\ {\small[ph cm $^{-2}$ s$^{-1}$]}} & $<$0.03 & 0.25 $\, ^{+0.03}_{-0.05}$   \\
\noalign {\smallskip}
& {\tt gau$_4$}  & & \\
\multicolumn{2}{l}{ \egau\ {\small[keV]}}               & 1.022 (f) & 1.022 (f)  \\
\multicolumn{2}{l}{ $\sigma$ {\small[keV]}}               &  0.01$^{}_{-0.01}$ & 0.1 $\, _{-0.01}$ \\
\multicolumn{2}{l}{ \kgau\ {\small[ph cm $^{-2}$ s$^{-1}$]}} & 0.11 $\pm$ 0.03 & 0.27 $\pm$ 0.03  \\
\noalign {\smallskip}
& {\tt gau$_5$}  & & \\
\multicolumn{2}{l}{ \egau\ {\small[keV]}}               & 2.28 $\pm$ 0.02 & 2.29 $\pm$ 0.03  \\
\multicolumn{2}{l}{ $\sigma$ {\small[keV]}}               &  0 (f) &  0 (f) \\
\multicolumn{2}{l}{ \kgau\ {\small[ph cm $^{-2}$ s$^{-1}$]}} & 0.039 $\pm$ 0.008  & 0.008 $\pm$ 0.004 \\
%\noalign {\smallskip}
%\noalign {\smallskip\hrule\smallskip}
%\multicolumn{2}{l}{{\flux}} &  & \\
%\multicolumn{2}{l}{{\small (10$^{-10}$ \ergcms)}} &  & \\
\noalign {\smallskip\hrule\smallskip}
        \multicolumn{2}{l}{\rchisq} & 1.27 & 1.85 \\
        \multicolumn{2}{l}{d.o.f.} & 274 &  237  \\
        \multicolumn{2}{l}{Exposure EPIC pn [ks]} & 0.69 &  0.64  \\
\noalign {\smallskip}
\multicolumn{2}{l}{L$_{0.5-200~keV}$ {\small [10$^{37}$ erg s$^{-1}$]}} & 7.68 & 3.95 \\
\noalign{\smallskip\hrule\smallskip}
\label{tab:bestfit}
\end{tabular}

\end{center}
\end{table}

Next, we added a blurred reflection component ({\tt kdblur*pexriv} in
XSPEC) to model 1. This component calculates the Compton-reflected
continuum but not the Fe \ka\ line. Thus we continued using the
relativistic line of the previous fits. The fits do not improve
significantly compared to the fits without reflection and the amount
of reflection is unconstrained. Further, the reflection component
cannot account for the excess observed at 1~keV mainly during
Obs~2. Finally, we substituted the power-law and {\tt pexriv}
components by the constant density ionized disk reflection model ({\tt
reflion.mod} component in XSPEC). This model measures the relative
strengths of the directly observed and reflected flux, the accretion
disk ionization parameter \xil\ and the photon index of the
illuminating power-law flux. Fe \ka\ line emission and line broadening
due to Comptonization in an ionized disk surface layer are included in
this model. The fits do not improve significantly compared to the fits
without reflection and the strong relativistic lines present in both
observations cannot be accounted for by this component. However we
note that, if included in addition to the relativistic line and the
power-law component, the {\tt reflion} component can account for the
excess at 1~keV in Obs~2, and the fit is improved significantly with a
final \rchisq\ of 1.70 for 234 d.o.f.. The reflection is ionized with
\logxi = 3.4 and a overabundance of Fe of 2.8 with respect to the
solar value is found. The blurring of the reflection component
indicates a different area of reflection with respect to the
relativistic line. In the case of reflection, the values of the inner
radius, the emissivity and the inclination are 1.86 GM/c$^2$, 10 and
75$\degmark$ respectively. No acceptable fit is found when the parameters
of the relativistic blurring are linked for the reflection component
and the relativistic line.

We conclude that the best-fit to the EPIC pn, JEM-X and
ISGRI spectra is model~1. The continuum can be well described by a
power law with a photon index, $\Gamma$, of $\sim$2.2-3.1
and a disk-blackbody with a temperature, \ktbb, of
$\sim$1.3~keV. The contribution of the power law to the total
2--10~keV luminosity is $\approxlt$4\% and $\approxlt$10\% for
Obs~1 and Obs~2, respectively. The hydrogen column density,
\nhabs, is 0.7--0.8~\ttnh. We can compare these values to
those obtained from  {\it Swift} observations \citep[see Tables 4
and 5 of ][]{1655:brocksopp06mnras}. The {\it Swift} observation
00030009005 was taken $\sim$ 8 hours later than XMM-Newton Obs~1.
It shows a similar value of \nhabs\
(0.69$^{+0.01}_{-0.02}$~\ttnh), a significantly smaller power-law
index (\phind\ = 1.7 $\pm$ 0.1) and a similar temperature for the
disk (\ktbb\ = 1.36 $\pm$ 0.02).
The
{\it Swift} observation 00030009008 was simultaneous with the
XMM-Newton Obs~2. It shows similar values of \nhabs\
(0.78 $\pm$ 0.03~\ttnh) and power-law
index (\phind\ = 3.3 $\pm$ 0.2) and a higher disk temperature
(\ktbb\ = 1.56 $^{+0.01}_{-0.02}$). The absence of a photo-ionized
absorber and relativistic Fe K emission in the modeling of the
{\it Swift} spectra probably contributes to these differences.

A relativistic emission feature centered on 6.60--6.80~keV is
detected in both XMM-Newton observations at a confidence level
$>$8$\sigma$, as determined by an F-test \footnote[2] {We note that
we use the F-test only as an indication for the significance of the feature.
For more realistic estimations, simulations are required \citep[see][]{protassov02apj}
.}. We obtained better fits
modeling the broadening of the line with a {\tt laor2}
component compared to the {\tt laor} component and to the Gaussian
broadening for both observations. The inclination of the {\tt
laor2} component is defined as the angle between the line of
sight and the rotation axis of the disk and was allowed to vary
between 50 and 90$\degmark$, since the inclination of the disk has
been determined to be 70$\degmark$ for \gro. The inner radius for
the emitting region has a value of 1.4--1.5 in units of
G$M$/c$^2$, indicating that the line is emitted very close to the
horizon of a Kerr black hole. The outer radius has been left fixed
to the initial value of 400 during the fits. The emissivity
profile is best described by a broken power-law, with {\it q$_1$}
= 9--10 within the break radius {\it r$_{{\rm break}}$} = 3.6--3.1
and {\it q$_2$} = 3.5--3.0 at larger radii. The steep emissivity
indicates that the disk is illuminated mostly in its very inner
regions supporting the idea that the primary source is very close
to the black hole. It is not possible to fit the emission line
unless we allow a very large emissivity. Comparable emissivity
indices ($\sim$7) have been observed from a number of Seyfert
galaxies \citep[e.g.,][]{0707:fabian04mnras, 0419:fabian05mnras}
and are indicative of a primary source very close to the black
hole.

\begin{figure}[ht!]
\hspace{-1.1cm}
\centerline{\includegraphics[width=0.5\textwidth]{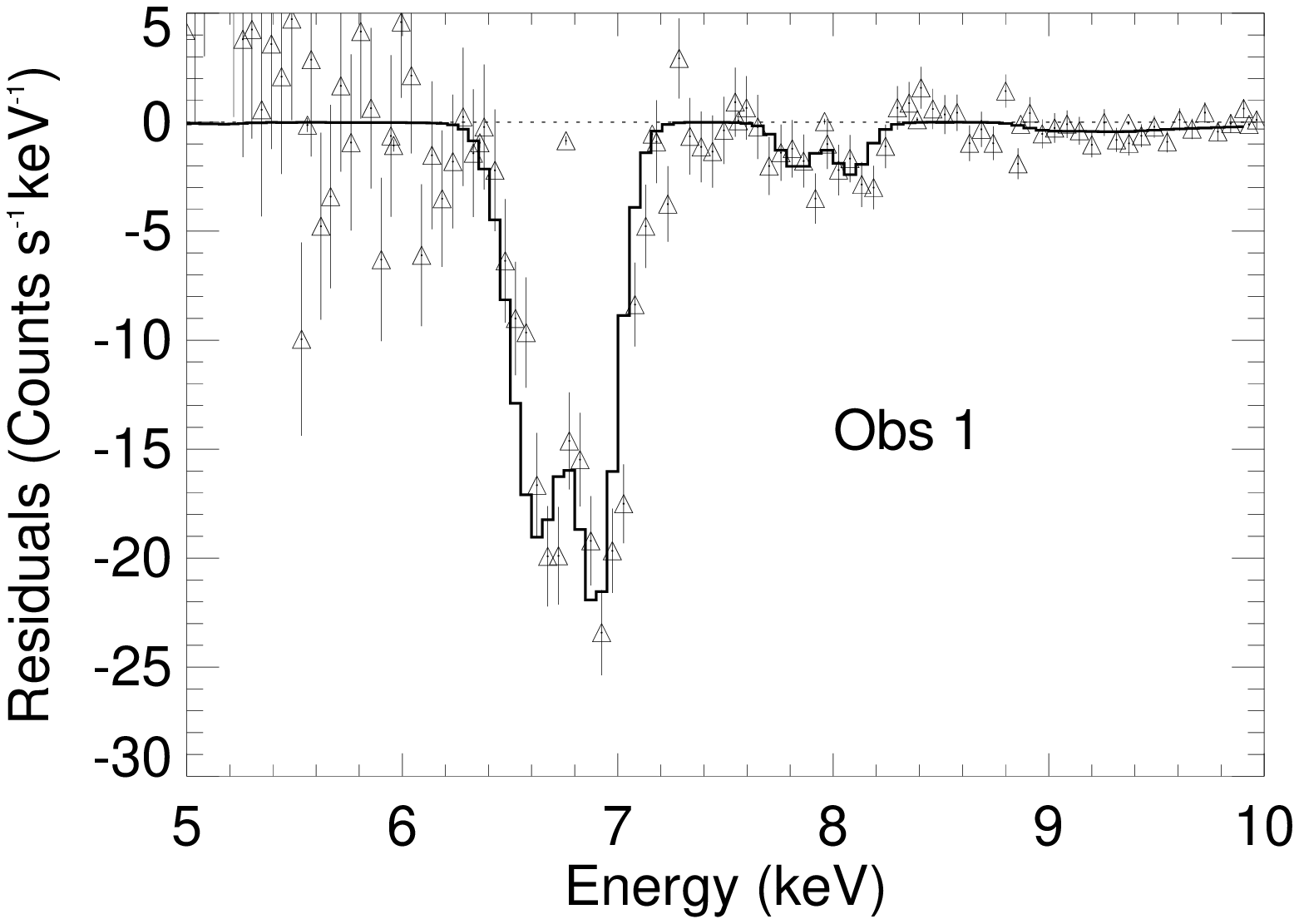}}
\hspace{-1.1cm}
\centerline{\includegraphics[width=0.5\textwidth]{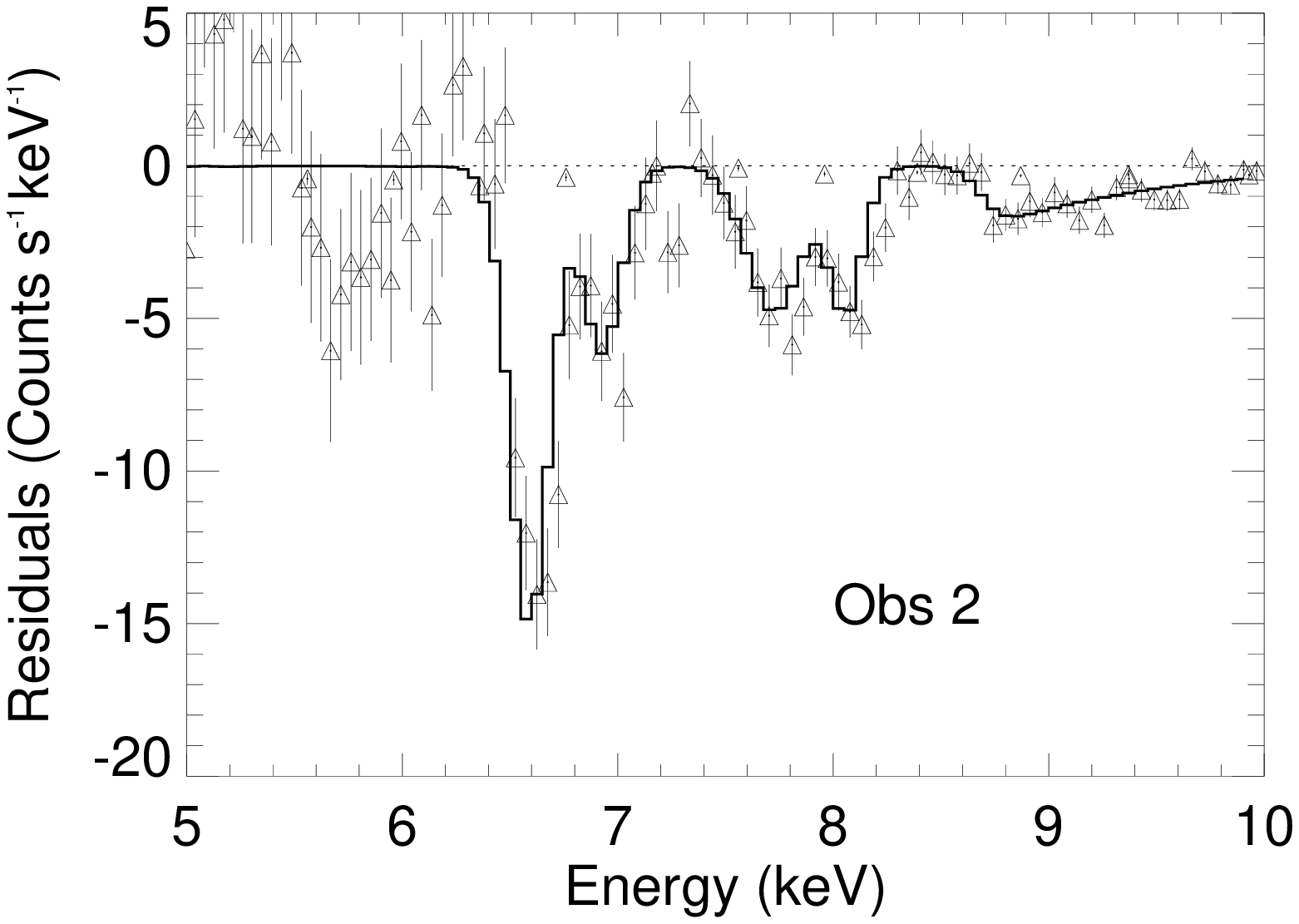}}
\caption{\src\ 5--10 keV EPIC-pn spectral residuals from the
best-fit {\tt tbabs*edge*edge*(dbb+po+gau$_6$+gau$_7$+gau$_8$+gau$_9$) + tbabs*(laor2+gau$_1$+gau$_2$+gau$_3$+gau$_4$) +gau$_5$} model
when the normalizations of the Gaussian absorption features {\tt gau$_{6-9}$} and the edges are
set to zero.} \label{fig:xspec-lines}
\end{figure}

We measured the \ew\ of the \fetfive\ and \fetsix\ absorption
features and edges produced by the warm absorber by removing the
warm absorber component and adding individually Gaussian
absorption features and edges to the model. The results are shown
in Table~\ref{tab:xspec-lines} and a detail of the absorption
features in Fig.~\ref{fig:xspec-lines}.
During Obs~1, the \ew\ of the \fetfive\ absorption feature is
smaller than that of the \fetsix\ \lya\ feature, whilst the
opposite occurs for Obs~2, indicating a less ionized absorber.
For Obs~2, the \ews\ are larger for the \lyb\ compared to the \lya\ features. This indicates that the \fetfive\ \lyb\ feature is rather a blend of \fetfive\ \lyb\ and \nitseven, while the \fetsix\ \lyb\ feature is a blend of \fetsix\ \lyb\ and \niteight.

\begin{table}[ht!]
\begin{center}
\caption[]{Best-fit of the most prominent EPIC pn absorption
features with negative Gaussian profiles for the two observations
of \gro\ using XSPEC. The $\sigma$ of the narrow Gaussian
absorption features are constrained to be $<$0.1~keV.}
\begin{tabular}{lc@{\extracolsep{0.15cm}}c@{\extracolsep{0.15cm}}c@{\extracolsep{0.15cm}}c@{\extracolsep{0.15cm}}}
\hline
%\noalign{\smallskip}
\hline
%\noalign{\smallskip}
 & \multicolumn{2}{c}{Obs 1} & \multicolumn{2}{c}{Obs 2} \\
\hline\noalign{\smallskip}
Lines & & & & \\
\noalign{\smallskip}
 & \multicolumn{2}{l}{\fetfive} & \multicolumn{2}{l}{\fetfive} \\
 & \ka\ & \kb\ & \ka\ & \kb\ \\
\egau\ {\small[keV]}  & 6.68 $\pm$ 0.03 & 7.88$\, ^{+0.16}_{-0.11}$  & 6.64 $\pm$ 0.02 & 7.78 $\pm$ 0.05 \\
\sig\ {\small[keV]}  & 0.1$\, _{-0.06}$ & 0 $\, ^{+0.1}$ & 0.06 $\pm$ 0.03 & 0.1$\, _{-0.07}$ \\
\ew\ {\small[eV]}  & 33 $\pm$ 9 & 11 $\pm$ 8 & 32 $\pm$ 6 & 36 $\pm$ 15 \\
\noalign{\smallskip}
&  \multicolumn{2}{l}{\fetsix} & \multicolumn{2}{l}{\fetsix} \\
& \ka\ & \kb\ & \ka\ & \kb\ \\
\egau\ {\small[keV]} & 6.96 $\pm$ 0.02 & 8.13$\, ^{+0.06}_{-0.20}$ & 6.98 $\pm$ 0.04 & 8.12 $\, ^{+0.02}_{-0.03}$ \\
\sig\ {\small[keV]} & 0.06 $\pm$ 0.03 & 0 $\, ^{+0.1}$ & 0 $\, ^{+0.1}$ & $<$0.07 \\
\ew\ {\small[eV]} & 39 $\pm$ 11 & 8 $\pm$ 4 & 18$\, ^{+6}_{-9}$ & 42$\, ^{+14}_{-28}$ \\
\hline\noalign{\smallskip}
\noalign{\smallskip}
Edges & & & & \\
\noalign{\smallskip}
 & \fetfive\ & & \fetfive\ & \\
\eedge\ {\small[keV]} & 8.96 $\pm$ 0.40 & & 8.74$\, ^{+0.03}_{-0.10}$ & \\
$\tau$  & $<$0.06 & & 0.18 $\pm$ 0.02 & \\
\noalign{\smallskip}
 & \fetsix\ & & \fetsix\ & \\
\eedge\ {\small[keV]} & 9.278 (f) & & 9.278 (f) & \\
$\tau$  & $<$0.05 & & $<$0.02 & \\
\noalign{\smallskip\hrule\smallskip}
\label{tab:xspec-lines}
\end{tabular}

\end{center}
\end{table}

\subsubsection{RGS spectral analysis}
\label{sect:rgs}
Next, we examined the 30 RGS1 spectra (first and second order taken
sequentially between 0.3 and 1.9~keV). The RGS2 spectra were
strongly piled-up and were therefore not used in the fits (see
Sect.~\ref{sec:reduction}). We note that the joint analysis of RGS
and pn data was not possible due to the presence of strong emission
features in the pn below 3~keV which were not detected in the
overlapping energy range of the RGS (see Sect.~\ref{sec:pn-isgri}) and
to the presence of detailed structure at the interstellar O~K and Fe~L edges 
in the RGS (see below) which was not resolved in the pn. We could fit the RGS1 
spectra of both observations with a continuum consisting of a power law
modified by photo-electric absorption from neutral material (\rchisq\
of 1.60 and 1.67 for 2183 and 2167 d.o.f.  for Obs~1 and Obs~2,
respectively). The RGS spectra show structured residuals near the
O edge at $\sim$0.54~keV. Around this energy there is a significant
change ($\sim$25\%) in the instrument efficiency, which may not be
fully accounted for in the data processing. Thus, the residuals could
be due either to a calibration uncertainties or have an astrophysical
origin. We checked for the second possibility by studying in detail
the region around 0.54~keV. We searched for the signature of O
absorption in the interstellar medium similar to that observed by
\citet{cygx2:takei02apj} and \citet{cygx2:costantini05aa} from
Cyg\,X-2 and by \citet{juett04apj} from a number of sources.

\begin{figure}[ht!]
\hspace{-1.1cm}
\centerline{\includegraphics[width=0.5\textwidth]{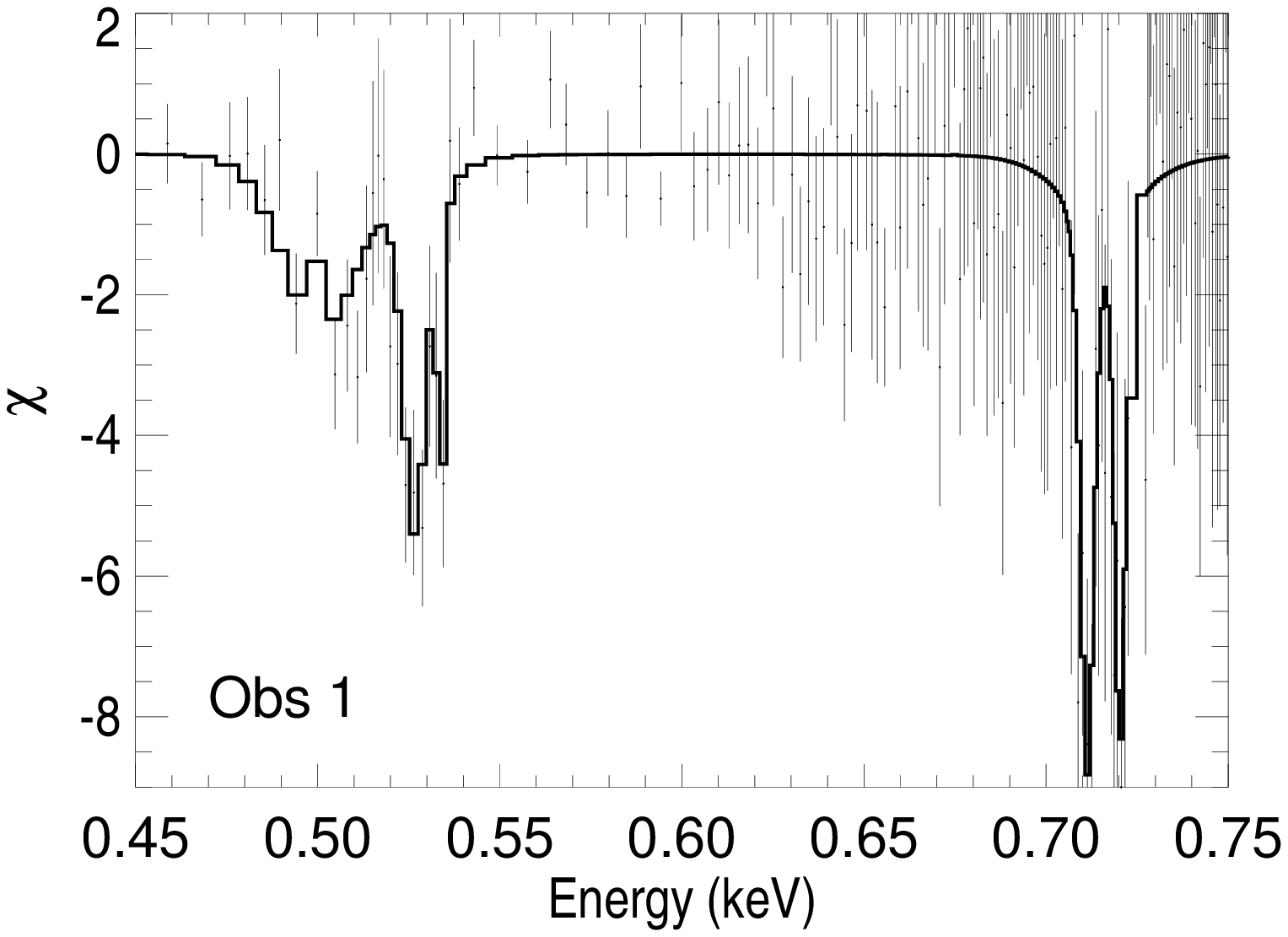}}
\hspace{-1.1cm}
\centerline{\includegraphics[width=0.5\textwidth]{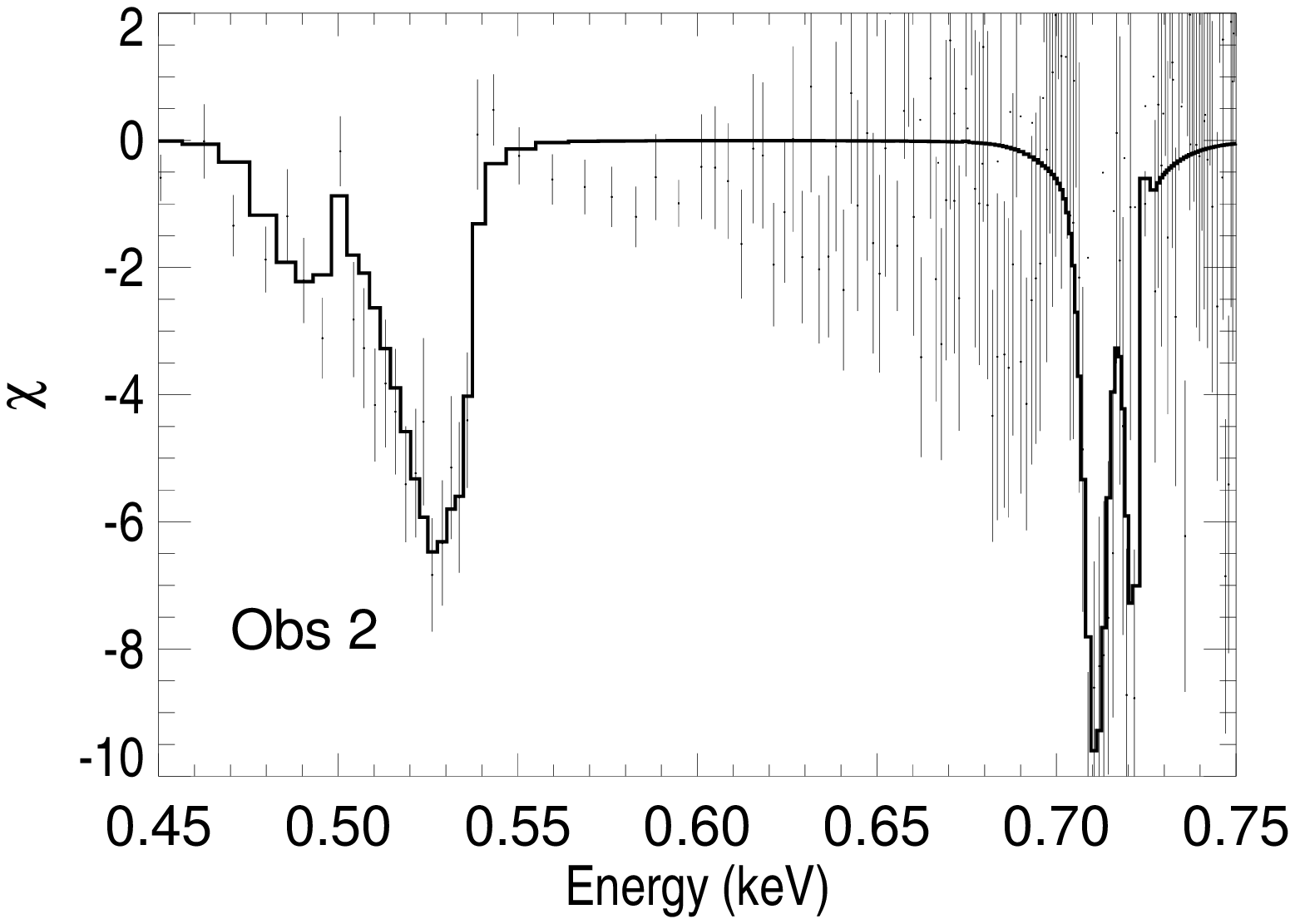}}
\caption{\src\ 0.45--0.75 keV RGS spectral residuals from the
best-fit model ({\tt tbabs*(pl+gau+gau+gau+gau+gau)} for Obs~1 and {\tt tbabs*warmabs*(pl+gau+gau+gau+gau+gau)} for Obs~2)
when the normalizations of the Gaussian absorption features are
set to zero.} \label{fig:rgs-ism}
\end{figure}

We included narrow Gaussian absorption features at energies of
0.498, 0.527 and 0.534~keV for Obs~1 and 0.488, 0.511 and
0.528~keV for Obs~2 (see Fig.~\ref{fig:rgs-ism}). This improves
the fit significantly and we identify the common feature at
0.528~keV with absorption from \oone\ and the feature at 0.534~keV
in Obs~2 with absorption from \otwo. We can not identify the other
absorption features found without assigning them extremely high
velocity shifts. However, we note that, if these features are of
interstellar origin, such velocity shifts are not expected for
galactic sources. A more plausible explanation is that at the low
detected count rates, data binning may cause several lines to
blend (see the extreme blending of the lines in Obs~2 compared to
Obs~1 in Fig.~\ref{fig:rgs-ism}). Thus, the energy of the detected
lines should be regarded rather as an average of the various O
lines present \citep[see e.g.,][]{juett04apj}. In addition, two
absorption features at 0.711~keV and 0.720~keV are evident in the
first and second order RGS spectra of the two observations. We
attribute these features to absorption by interstellar Fe in the 
L3 (2p$_{3/2}$) and L2 (2p$_{1/2}$) edges, 
respectively. Including narrow Gaussian absorption
features to model these residuals reduces the \rchisq\ to 1.53 for
2168 d.o.f for Obs~1 and to 1.53 for 2152 d.o.f. for Obs~2. The
energies and equivalent widths of all the detected features are
listed in Table~\ref{tab:rgs-lines}. The residuals of the
best-fits to the RGS spectra when such lines are excluded are
shown in Fig.~\ref{fig:rgs-ism}.

\begin{figure}[ht!]
\centerline{\includegraphics[width=0.6\textwidth]{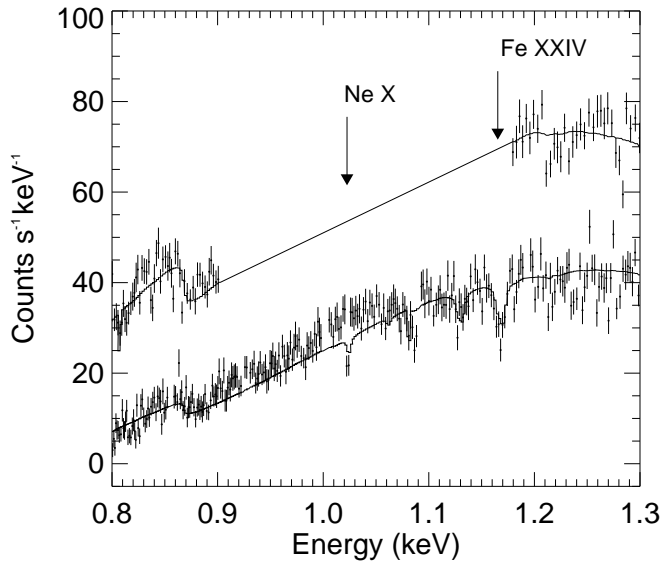}}
\caption{First and second order RGS1 \gro\ spectra of Obs~2.
Two narrow absorption features are detected at 1.023 $\pm$ 0.002 and 1.168 $\pm$ 0.002~keV and
identified with \neten\ and \fetfour, respectively.}\label{fig:rgs}
\end{figure}

The Obs~1 spectra do not show any apparent features except for the
ones previously mentioned and including absorption from a
photo-ionized plasma in the model does not improve the quality of the
fit. In contrast, the Obs~2 spectra show a number of low intensity
absorption features superposed to the continuum, in addition to the
ones attributed to absorption in the interstellar medium. To account
for these features absorption from a photo-ionized plasma (component
\warmabs\ in XSPEC) was added to the model. An ionized absorber with
\nhwarmabs\ = (3.2 $\pm$ 1.6) $\times 10^{22}$~atom~cm$^{-2}$,
\logxi\ = 3.3 $\pm$ 0.1, \sigmav\ = 160 $\pm$ 60 \kms\ and $v$
= $-$540 $\pm$ 120 \kms\ fits well the absorption features and the
final \rchisq\ is 1.45 for 2148 d.o.f (note that we keep here the
parameters of the continuum fixed to the values of the previous fit).
Fig.~\ref{fig:rgs} shows the
best-fit model to the RGS spectra of Obs~2.
The lines are significantly
blue-shifted, indicating that the absorbing matter is an outflow along
our line of sight. The blue-shift detected in the RGS spectra which
have an energy resolution of $\sim$300 at 1~keV, is too small to be
detected with the pn camera, which has an energy resolution of
$\sim$45 at 6.4~keV and is only sensitive to shifts
$\approxgt$1000~\kms. Two narrow absorption features are detected at 1.023 $\pm$ 0.002
and 1.168 $\pm$ 0.002~keV with \ews\ of $<$1~eV and 4.3 $\pm$ 1.2~eV and are identified
with \neten\ and \fetfour, respectively.
The column density and photo-ionization parameter of the absorber are
consistent within the uncertainties with the absorber fit to the pn
spectrum, showing that the same outflow is able to produce the
\neten\ and \fetfour\ features at $\sim$1~keV and the highly ionized
\fetfive\ and \fetsix\ features at $\sim$7~keV.

We next fitted the abundances of O, Ne and Fe, which were
previously fixed at solar values. We obtained values of $<$0.8,
5$^{+3}_{-4}$ and 4 $\pm$ 3 for O, Ne and Fe, respectively, with
respect to solar values. The parameters of the ionized absorber
are fixed during the fit of the abundances and the fit is
improved, but not significantly, with a final \rchisq\ of 1.44 for
2142 d.o.f.
Thus, we find no evidence for a significant overabundance of
any element, with respect to solar values. Unusual Ne/O abundance
ratios have been reported for several ultracompact X-ray binaries
and have been attributed to a donor rich in Ne \citep[e.g,
][]{juett03apj}. Variations in the Ne/O abundance ratio may be
related to ionization effects due to continuum spectral variations
\citep[see ][and references therein]{juett05apj}. We note that
using a self-consistent model for the photo-ionized absorber, no 
significant Ne/O overabundance is detected in the RGS spectra.

\begin{table*}[ht!]
\begin{center}
\caption[]{Best-fit of the most prominent absorption features with
negative Gaussian profiles for the two \gro\ observations. The
fits were performed using XSPEC on the RGS spectra.
f indicates that a parameter was fixed during the
fitting. The $\sigma$ of the narrow Gaussian absorption features
is constrained to be $<$0.01~keV.}
\begin{tabular}{cccccc}
\hline
%\noalign{\smallskip}
\hline
%\noalign{\smallskip}
\multicolumn{3}{c}{Obs 1} & \multicolumn{3}{c}{Obs 2} \\
\hline\noalign{\smallskip}
\noalign{\smallskip}
\egau\ {\small[keV]} & \sig\ {\small[keV]}  & \ew\ {\small[eV]}  & \egau\ {\small[keV]} & \sig\ {\small[keV]}  & \ew\ {\small[eV]} \\
\noalign{\smallskip}
\noalign{\smallskip}
0.498 $\, ^{+0.011}_{-0.003}$ & 0.010 $\, _{-0.002}$ & 14 $\, ^{+5}_{-3}$ & 0.488$\, ^{+0.004}_{-0.008}$ & 0.010$\, _{-0.002}$ & 19 $\pm$ 5\\ 
0.527 $\pm$ 0.003 & 0.003 $\pm$ 0.001 & 9 $\, ^{+4}_{-3}$ & 0.511$\, ^{+0.003}_{-0.006}$ & 0.006 $\pm$ 0.004 & 7$\, ^{+6}_{-3}$\\
0.534 $\pm$ 0.004 & $<$0.01 & $<$6 & 0.528$\, ^{+0.002}_{-0.004}$ & 0.009$\, ^{+0.007}_{-0.002}$ & 23$\, ^{+7}_{-3}$ \\
0.711 $\pm$ 0.001 & $<$0.003 & 3 $\pm$ 1 & 0.711 $\pm$ 0.002 & $<$0.004 & 5 $\pm$ 3\\
0.720 $\pm$ 0.001 & $<$0.003 & 2 $\pm$ 1 & 0.721$\, ^{+0.002}_{-0.005}$ & $<$0.006 & 2$\, ^{+5}_{-1}$\\
\noalign{\smallskip}
\hline\noalign{\smallskip}
\noalign{\smallskip}
& & & 1.023 $\pm$ 0.002 & $<$0.002 & $<$1 \\
& & & 1.168 $\pm$ 0.002 & 0.003 $\pm$ 0.002 & 4.3 $\pm$ 1.2 \\
%& & & 1.553$^{+0.005}_{-0.002}$ & & 1.4 $\pm$ 1.0 \\ 
\noalign{\smallskip}
 
\noalign{\smallskip\hrule\smallskip}
\label{tab:rgs-lines}
\end{tabular}

\end{center}
\end{table*}

\subsection{Time-resolved spectral fits during Obs~1}

As Fig.~\ref{fig:lc} shows, both the flux and hardness ratio
of \gro\ increased during Obs~1. To study the variations of the
highly-ionized absorber with the increase in flux, we divided the
light curve of Obs~1 into two segments. 
We extracted two RGS spectra, 
set 1 comprises from the beginning till the middle of the
observation and set 2 from the middle till the end, since the
middle ``dip'' interval had very few counts. 

Both sets can be fit with the same model as for the
complete observation which is a continuum consisting of a power
law modified by photo-electric absorption from neutral material
(\rchisq\ of 1.41 and 1.44 for 1865 and 1971 d.o.f. for first and
second sets, respectively). The power-law index is the same for
both observations (\phind = 1.38). Similarly, the column density of the
neutral absorption, \nh, does not change significantly from the first (\nh =
(7.60 $\pm\, 0.05)\, \times\, $10$^{21}$ cm$^{-2}$) to the third
set (\nh = (7.7 $\pm\, 0.1)\, \times\, $10$^{21}$ cm$^{-2}$).
Including absorption by an ionized slab into the model does not
improve the fit significantly for either set.

\section{Discussion}

We have analysed two observations of \gro\ during its 2005
outburst. During Obs~1, which occurs close to the maximum of the
first peak of the outburst, the 0.5--10~keV luminosity is still
increasing, and it is clearly in the high-soft state, showing an
X-ray emission dominated by the thermal emission (the 2--10~keV
luminosity is $\approxgt$96\%), probably from the optically thick
accretion disk, and a weak, steep hard X-ray power law. Obs~2 is
close to the minimum of the first peak of the outburst, nine days
later, and shows a lower, less-variable luminosity. However the
characteristics of the source still resemble those of a high-soft
state. The absence of quasi-periodic oscillations (QPOs) in the
RXTE data from 2005 March 18 between 19:41 and 21:41~UTC and from
2005 March 27 between 07:46 and 11:56~UTC, simultaneous to the
analyzed XMM-Newton and INTEGRAL first and second observations,
respectively, indicate that \gro\ is not in the "steep power law"
state, as defined by \citet{Mcclintock06}, but rather in the
high-soft state, where the power law is weak (less than 10\% of the
total 2--10~keV luminosity) and has indices between 2.1 and 4.8,
and QPOs are absent in the power spectrum.

A highly-ionized absorber is observed superposed on the continuum
in both observations. The absorber is less ionized and has a
smaller column density during Obs~2, where blue-shifted absorption
features of \neten\ and \fetfour\ are detected in the 0.3--1.9~keV
RGS spectrum, in addition to the EPIC pn \fetfive\ and \fetsix\
features. A range of absorbers with different temperature or
density is not required to explain the features at $\sim$7~keV and
below 2~keV. Presumably, the ionized absorber is an outflow in
both observations, but we can only give upper limits (the EPIC pn
energy resolution is $\sim$1000~\kms at 6~keV) to the outflow
velocity in Obs~1 due to the absence of features in the RGS
spectrum. The relation of luminosity of the X-ray source with the
accretion rate has been extensively discussed \citep[e.g.,
][]{1655:hynes98mnras}. In the case that the lower luminosity
during Obs~2 indicated that the accretion rate was smaller, the
absorber column could be related to the accretion rate. Further,
the value of the ionization parameter of the absorber is directly
related to the luminosity ($\xi$ = L/n$_e$r$^2$). Therefore a
lower degree of ionization is expected in Obs~2, which has a lower
luminosity, as detected.

Narrow X-ray absorption lines from highly ionized Fe were first
detected in superluminal jet sources \citep{1655:ueda98apj,
1655:yamaoka01pasj, 1915:kotani00apj, 1915:lee02apj} and subsequently
in an increasing number of low-mass X-ray binaries observed by
XMM-Newton \citep[][ and references therein]{gx13:ueda01apjl,
ionabs:diaz06aa}. Recently, {\it Chandra} HETGS observations of the
black hole candidates \threethreenine, \xte\ and \seventeen\
\citep{gx339:miller04apj,1743:miller04apj} have revealed the presence
of variable, blue-shifted, highly-ionized absorption features which
are interpreted as evidence for outflows.

The detection of an outflowing highly-ionized absorber
in \gro\ confirms that the outflowing winds present in AGNs are
also a common feature of microquasars. 
In contrast, no blueshifts
have been detected in any of the highly-ionized absorbers present
in dipping LMXBs \citep{ionabs:diaz06aa}. However, these results
are all obtained with the EPIC pn, which has a factor $\sim$4
poorer resolution than the HETGS at 6~keV limiting the sensitivity
to shifts $\approxgt$1000~\kms. Therefore, we calculated the upper
limits for the velocity shift in \mxb, the only LMXB which shows
absorption features of \oeight\ and \neten\ in the RGS. We obtain
a velocity shift of $-$215$^{+245}_{-270}$ \kms\ for the absorber,
indicating only a marginal blueshift. Attempts have been made to
relate the ionized absorption with jet emission, first due to the
discovery of absorbers in microquasars preceding the discovery in
LMXBs \citep[e.g.,][]{1655:ueda98apj}, and later due to the
difference in velocity shift between both kinds of systems, i.e.
an outflowing wind in microquasars versus an static atmosphere in
LMXBs \citep{1743:miller04apj}.

The detection of ionized absorbers in many highly-inclined systems
points to a common origin for the absorbers in LMXBs and microquasars,
and not related to the jets of the latter. Further, with the data
available up to now we can not exclude a velocity shift in the
absorbers of LMXBs, which marks a difference between the absorbers in
galactic black holes and LMXBs. Measurements of shifts of low ionized
features with the RGS on-board XMM-Newton or HETGS on-board {\it
Chandra} are necessary in order to gain insight into the distribution
and origin of the ionized absorbers.  The detection of a different
column density and ionization state of the absorber in different
phases of the outburst establishes a clear link between the
variability of the warm absorber and the central accretion
engine. This link is confirmed when comparing the highly-ionized
absorption lines detected by ASCA \citep{1655:ueda98apj} with the ones shown in
Table~\ref{tab:xspec-lines}: an increase in the flux from the
``low-state'' ASCA observation to XMM-Newton Obs~2 and Obs~1 and
finally to the ``high-state'' ASCA observation is accompanied by an
increase of the ionization parameter, \xil. While only \fetfive\
features are detected in the ``low-state'' ASCA observation, both
\fetfive\ and \fetsix\ are seen in the XMM observations and only a
\fetsix\ feature remains in the ``high-state'' ASCA observation. The long
dynamical time scales make it difficult to perform studies of
variability of the warm absorber with the central accretion engine in
AGNs. Thus future observations of galactic black holes may also
provide insights in accretion processes in AGNs.

A relativistically broadened Fe \ka\ emission line is detected in
both observations. The inner emission radius of the line is
smaller than 6R$_g$, indicating that \gro\ harbors a black hole
with significant spin. A further indication of the existence of a
spinning black hole in \gro\ is the presence of high-frequency
QPOs \citep{1655:strohmayer01apjl,1655:abramowicz01aal}. All the
parameters of the line, except the emissivity, are consistent with
the relativistically broadened Fe \ka\ line previously detected by
ASCA \citep{asca:miller04}. The energies of the lines, 6.60
and 6.80~keV, are consistent with ionized emission from \fetfive\
and lower ionization stages from Fe, such as \fetfour\ and
\fetthree\ for Obs~1, and from \fetfive\ and \fetsix\ for Obs~2.
Ionized Fe K emission is expected from a disk with such a high
temperature (kT $\sim$ 1.3~keV). The discrepancy in the
emissivity, $\sim$9-10 in our observations compared to $\sim$5.5
in ASCA observations, may be a consequence of the model chosen to
perform the fits. The spectra fit in this paper show highly
ionized absorption lines, which are modeled with a photo-ionized
absorber. The absorption is strong, \nhxabs = (5.2 $\pm$ 1.0) and
(1.5 $\pm$ 1.0)~\ttnh\ for Obs~1 and 2, respectively, and causes a
change in the continuum. The relativistic emission line occurs at
the same energy interval than the strongest absorption lines and
edges, and therefore an absent or deficient modeling of the
absorption (e.g. by including the lines and edges independently)
will vary the parameters of the emission feature. Further, the
ASCA fits include a smeared edge at 8~keV to simulate
reflection from the disk. This edge is not included in our fits.
Finally, the inclination in the ASCA fits (45$\, ^{+15}_{-5}$
degrees), though consistent within the errors, is lower than the
one in our fits (52 $\pm$ 1 and 63.2 $\pm$ 0.2 degrees). This may
explain partially the high emissivity of our fits ($q_1\sim$9-10),
since an increase in the emissivity produces the same effect as a
lower inclination in the relativistic line. The simultaneous
modeling of the absorption features and of the relativistic Fe K
emission line is crucial in order to obtain realistic values for
the parameters of the model. It has been argued
\citep[e.g.,][]{1650:done06mnras} that relativistic smearing can
be significantly reduced if there is also Fe K line absorption
from an outflowing disk wind. We find that {\it both} an extremely
relativistic emission line, confirming the existence of a spinning
black hole, and strong ionized absorption are present in the
outburst spectra of \gro.

The role of reflection and absorption in AGNs has been extensively
discussed \citep[e.g.,][]{agn:chevallier06aa}. The reflection models
used in our fits cannot account for the strong relativistic Fe K line
and do not improve significantly the fits. Further, they can not
explain either the strong excess at 1~keV of Obs~2, unless the
region of reflection is different to the region of emission of the
relativistic line, an extra power-law component is included and an
overabundance of 2.8 of Fe with respect to solar values is allowed. A
narrow emission line is observed at 0.6~keV in Obs~1.  Such line at
0.6~keV could be associated to emission from \oseven, if its
astrophysical origin is confirmed. However, its non-detection in the
RGS data may indicate that the line has an instrumental origin. The
feature at 1~keV, detected at high significance in Obs~2, and
only marginally in Obs~1, has been observed in other sources and its
origin is unclear. The fact that it is stronger in the less
luminous \gro\ observation indicates that it could have an
astrophysical origin. In that case, it could be related to a
combination of \nenine\ and \neten\ or the Fe L complex.

\section*{Note added after submission}

One month after submission of this paper to A\&A, a paper by
\citet{1655:miller06nat} reporting on a {\it Chandra} observation of
\gro\ performed on 1 April 2005, 5~days after Obs~2, was accepted for
publication. The HETGS spectrum shows 90 absorption lines significant
at the 5$\sigma$ significance level or higher, from which 76 are
identified with resonance lines expected from over 32 charge
states. The equivalent widths of the lines detected by {\it Chandra}
are a factor $\sim$1.5 larger than the ones of the corresponding lines
detected in XMM-Newton Obs~2. This indicates an increase in the column
density of the absorber from the XMM-Newton Obs~2 to the {\it Chandra}
observation. 
This change in the column
density and the difference in exposure times, 0.64~ks(1.8~ks) in the pn(RGS)
for Obs~2 versus 63.5~ks in the HETGS, explain the detection of
significantly fewer lines in the XMM-Newton observation.

%%%%%%%%%%%%%%%%%%%%%%%%%%%%%%%%%%%%%%%%%%%%%%%%%%%%

\begin{acknowledgements}
Based on observations obtained with XMM-Newton, an ESA science
mission with instruments and contributions directly funded by ESA
member states and the USA (NASA), and on observations with
INTEGRAL, an ESA project with instruments and science data centre
funded by ESA member states (especially the PI countries: Denmark,
France, Germany, Italy, Switzerland and Spain), Czech Republic and
Poland, and with the participation of Russia and USA. M. D{\'i}az
Trigo acknowledges an ESA Research Fellowship. We thank the RXTE
instrument teams at MIT and NASA/GSFC for providing the ASM
lightcurve. We thank the anonymous referee for helpful comments.
M. D{\'i}az Trigo thanks Tim Kallman for his help with the {\tt warmabs} model. 
\end{acknowledgements}

%------------------------------------
%       References
%------------------------------------

\bibliographystyle{aa}
\bibliography{mybib}

\end{document}